\documentclass[12pt,preprint]{aastex}

\usepackage{natbib}
\usepackage{amsfonts}
\usepackage{color}
\usepackage{ulem}
\usepackage{amsmath}
\usepackage{mathtools}
\newcommand{\rsun}{$\,R_{S}$}
\newcommand{\degrees}{$^{\circ}$}
\newcommand{\thetabn}{$\theta_{BN}$}

\shorttitle{Coronal Shocks and Energetic Particles}
\shortauthors{Kozarev K. A., Raymond J. C., Lobzin V. V., Hammer, M.}

\begin{document}


\title{Properties of a Coronal Shock Wave as A Driver of Early SEP Acceleration}

\author{K. A. Kozarev}
\affil{Smithsonian Astrophysical Observatory,
    Cambridge, MA 02138}
\email{kkozarev@cfa.harvard.edu}

\author{J. C. Raymond}
\affil{Harvard-Smithsonian Center for Astrophysics,
    Cambridge, MA 02138}

\author{V. V. Lobzin\altaffilmark{1}}
\affil{Learmonth Solar Observatory, Bureau of Meteorology, Exmouth, WA 6707, Australia}
\altaffiltext{1}{Also at School of Physics, University of Sydney, NSW 2006, Australia}
 
\author{M. Hammer}
\affil{Department of Physics, Cornell University, 
    Ithaca, NY 14853}
    




\begin{abstract}
Coronal mass ejections (CMEs) are thought to drive collisionless shocks in the solar corona, which in turn have been shown capable of accelerating solar energetic particles (SEPs) in minutes. It has been notoriously difficult to extract information about energetic particle spectra in the corona, due to lack of in-situ measurements. It is possible, however, to combine remote observations with data-driven models in order to deduce coronal shock properties relevant to the local acceleration of SEPs and their heliospheric connectivity to near-Earth space. We present such novel analysis applied to the May 11, 2011 CME event on the western solar limb, focusing on the evolution of the eruption-driven, dome-like shock wave observed by the Atmospheric Imaging Assembly (AIA) EUV telescopes on board the Solar Dynamics Observatory spacecraft. We analyze the shock evolution and estimate its strength using emission measure modeling. We apply a new method combining a geometric model of the shock front with a potential field source surface model to estimate time-dependent field-to-shock angles and heliospheric connectivity during shock passage in the low corona. We find that the shock was weak, with an initial speed of $\sim$450 km/s. It was initially mostly quasi-parallel, but significant portion of it turned quasi-perpendicular later in the event. There was good magnetic connectivity to near-Earth space towards the end of the event as observed by the AIA instrument. The methods used in this analysis hold a significant potential for early characterization of coronal shock waves and forecasting of SEP spectra based on remote observations.
\end{abstract}

\section{INTRODUCTION}
\label{s1}
The early phase of coronal mass ejections (CMEs) usually is quite impulsive, driven by loop expansion, filament eruption or both. Extreme ultraviolet (EUV) observations from the Atmospheric Imaging Assembly (AIA) instrument \citep{Lemen:2012} on the Solar Dynamics Observatory (SDO) spacecraft over the past several years have allowed an unprecedented new look into the low coronal evolution of such events with their superb temporal and spatial resolution. The 193 and 211-Angstrom bands of AIA have been shown capable of observing shock waves low in the corona \citep{Ma:2011, Kozarev:2011, Patsourakos:2010}. Such observations give detailed information about the shape, kinematics, and varying thickness of the waves. A host of observations in the last several years have revealed much about the properties of coronal bright fronts on the solar disk as well as off the limb \citep{Nitta:2013}, provided strong evidence that they are indeed wavelike \citep{Long:2011, Olmedo:2012}, and should be treated as truly three-dimensional phenomena, not just surface disturbances \citep{Veronig:2010}. Wave observations off the solar limb are best suited for studying the leading front evolution, as they allow for far more accurate determination of the global front shape, thickness, and time-dependent position. Such observations can also be used to estimate the density and temperature changes in the sheath behind the fronts via emission measure modeling, using the six EUV channels of AIA \citep{Kozarev:2011}. This is a unique opportunity to amass statistics on the properties of fast eruptive events, and to determine whether the shocks they drive can indeed accelerate particles to tens or hundreds of MeV in the short time they spend in the AIA field of view. To that end, we have begun a concentrated effort to develop a framework for semi-automated analysis of off-limb eastern and western events. Studying off-limb events removes much ambiguity regarding the radial origin of the optically thin EUV emission, which is important for better kinematics characterization.

EUV bright large-scale fronts are often accompanied by type II metric radio bursts, which are attributed to coronal shocks (see, e.g. \citet{Pick:2008}). A recent study showed for the first time through combined radio and EUV imaging the strong temporal and spatial co-occurrence of coronal waves and CME-driven shocks \citep{Carley:2013}. Coronal shocks, in turn, are often associated with solar energetic particle (SEP) events. However, to date little detailed work has been done on how, where, and when energetic particles get accelerated in shocks traveling through the corona. In a detailed global modeling study, \citet{Kozarev:2013} showed that the evolution of a CME and the shock it drove in the low and middle corona strongly affected the SEP fluxes, as a function of both time and position around the shock front. Our goal is to develop a framework for extracting the parameters that determine the efficiency with which shocks may accelerate particles very early in events and close to the Sun, using remote observations and observations-driven models. To present our methods, we present here an analysis of the initial phase of a CME on May 11, 2011, which originated near the northwest solar limb. It featured an erupting filament, which drove a large-scale dome-shaped bright front.

This paper is structured as follows: In Section~\ref{s2}, we describe the May 11, 2011 event, presenting remote EUV and metric radio, as well as 1 AU particle observations. In Section~\ref{s3}, we characterize the observed shock wave characteristics, important for SEP acceleration. For this, we use shock kinematics and emission measure (EM) analysis based on the AIA observations.  We introduce and apply a new technique, which combines AIA observations with potential field source surface modeling (PFSS) to estimate time-dependent shock-to-field angle \thetabn~and the portion of the coronal surface, to which shock-accelerated SEPs may have access during the initial event stages. Finally, we discuss the implications for particle acceleration and coronal connectivity, and summarize our results in Section~\ref{s4}.


\section{Observations of The May 11, 2011 Event}
\label{s2}
On May 11, 2011, the initial stage of an eruption event was visible in the AIA field of view (FOV) in the interval ~02:12-02:31 UT, and was associated with a B8.1 flare and an erupting filament from AR 11205 near the NW limb. It coincided with metric radio type II emission, and significant 2-50 MeV proton flux enhancement near Earth followed it. Below, we review the most relevant observations of the event.

\begin{figure}[htc]
\noindent\includegraphics[width=1.0\columnwidth]{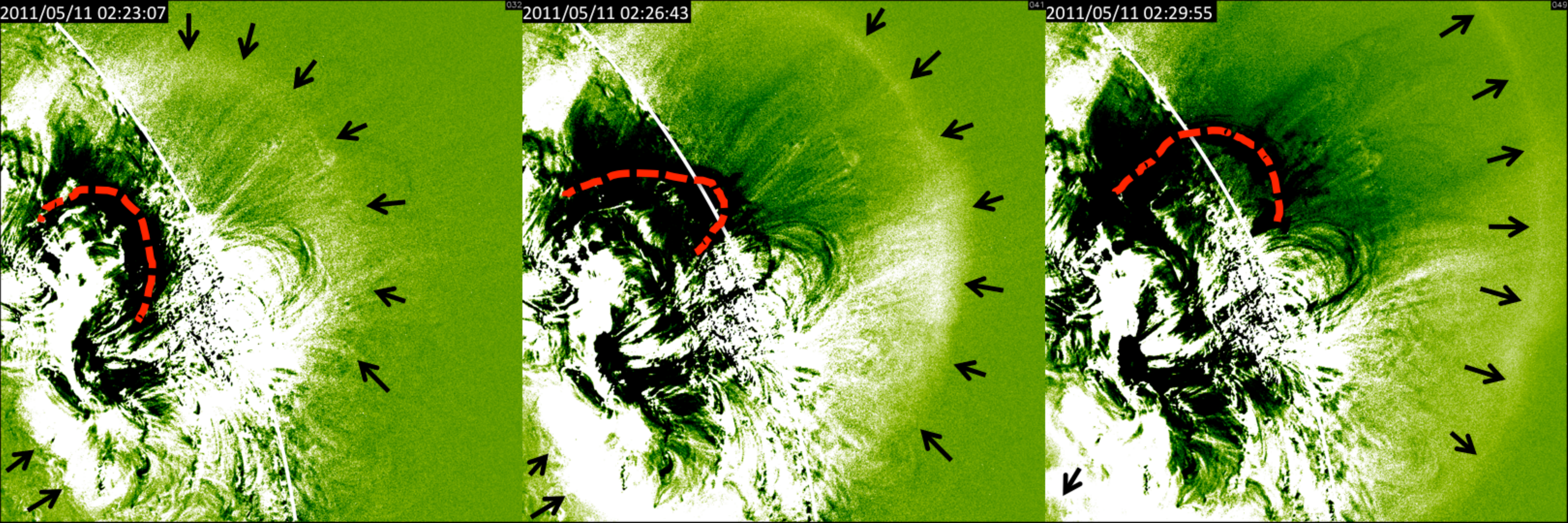}
\caption{Three AIA/193 channel base difference images showing the evolution of the May 11, 2011 off-limb wave. The dome-shaped wave left traces on the disk as well as off-limb. Driven by a large erupting filament (outlined with dashed red curves), it developed a thin front easily seen in these images, and outlined by arrows. A movie is available in the online version of this manuscript.}
\label{fig_wave}
\end{figure}

\subsection{EUV Observations}
We followed the dynamics of the erupting flux rope and bright front surrounding it in images from the EUV 193~$\AA$ channel of AIA (Fig. \ref{fig_wave} and accompanying movie). This channel, in our experience, shows the fronts most clearly. A shock is not resolved; instead, what is observed is a quasi-spherical sheath that exhibits a sharp increase in intensity at its front edge - the position of this increase is commonly taken as the shock front \citep{Vourlidas:2003, Ontiveros:2009, Kozarev:2011, Ma:2011}. Throughout this paper, we use the term Coronal Bright Front \citep[CBF]{Long:2011} to refer to the phenomenon that can be described as a `shock wave', `shock front', `bright front', defined as the anti-sunward edge of the shock sheath. The EUV image counts in that feature are much lower than the emission of the surrounding Sun, so for the purpose of studying its dynamics we have prepared 193~$\AA$ base difference images by subtracting a pre-event image taken at 02:10:19~UT from all consecutive frames (separated by approximately 12-seconds each). That was also the time when the relatively dark filament began to slowly expand. Shortly thereafter, enhanced emission in small local patches underneath it signified the onset of reconnection and possible Rayleigh-Taylor instabilities. The CBF becomes distinctly visible in the base difference images around 02:17:43 UT in the 193 channel. Interestingly, this coincides with the time when large-scale reconnection begins below the hitherto slowly expanding filament. \citet{Gopalswamy:2012} noted a similar relationship between the flux rope and coronal wave in the June 13, 2010 event - the shock sheath became distinctly visible at the time when the flux rope began intense lateral expansion.

\begin{figure}
\noindent\includegraphics[width=1.0\columnwidth]{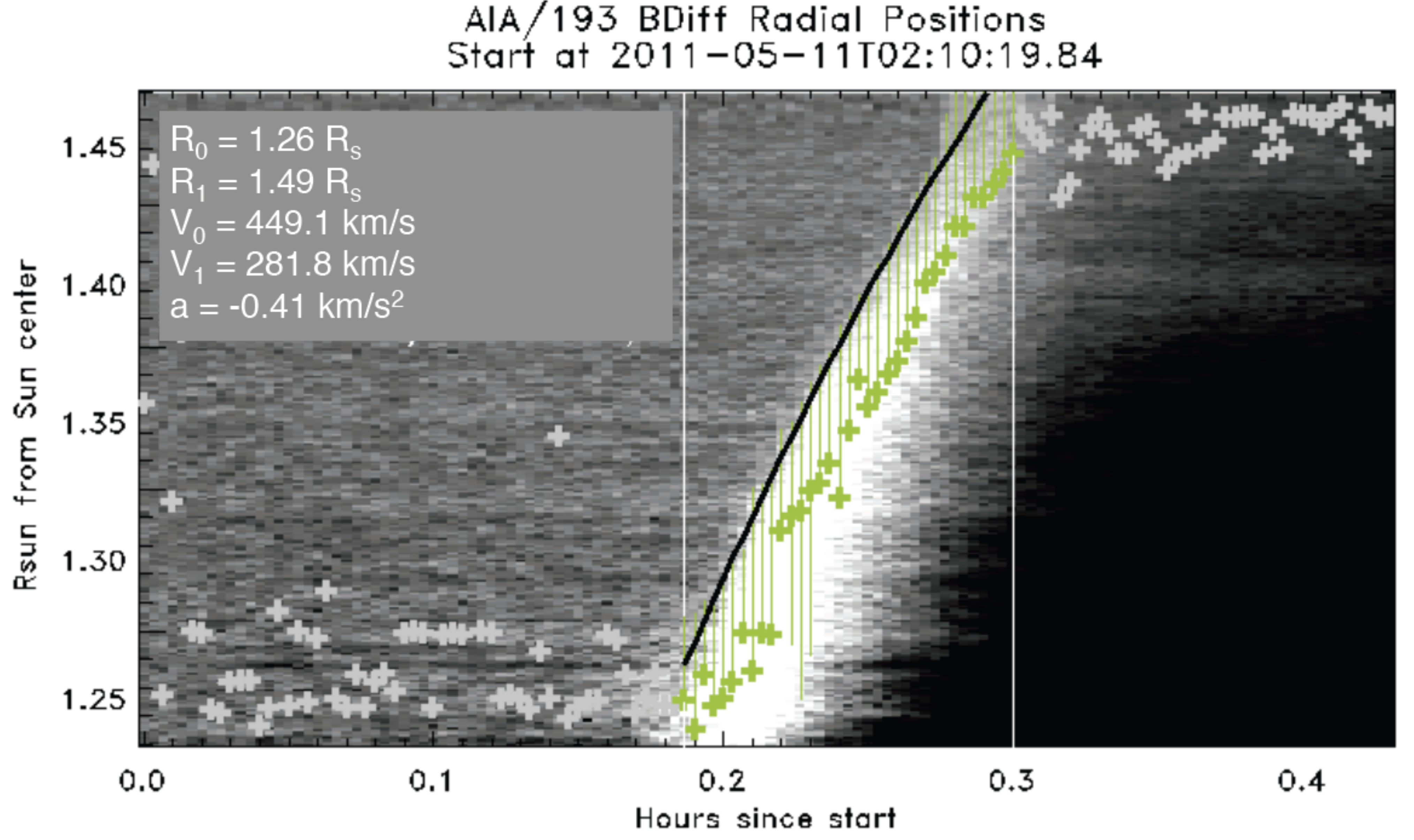}
\caption{The wave front positions were measured along a radial line through the CBF nose for each time step by an automatic algorithm, which finds local and global intensity peak positions. The wave front positions are based on the locations where intensity is 20\% of the peak values. The vertical white lines denote the times of start and end of the off-limb CBF in the AIA FOV.}
\label{fig_kinematics}
\end{figure}

We analyzed the radial evolution of the CBF by taking base-differenced intensity measurements along its nose over the event duration in the AIA FOV. These were then stacked, one measurement per time step, on a time-height plot (Fig. \ref{fig_kinematics}). The peak intensity at each time step was found using a local extrema algorithm. We determined the starting time of the off-limb portion of the CBF by eye, and the ending position when the CBF exited the AIA FOV (vertical white lines in Fig. \ref{fig_kinematics}). We defined the CBF position as the location where the intensity value is 20\% of the peak for that time step. The measured peaks are denoted with green `+' symbols in the figure. We fitted second-order polynomials to the front positions in order to obtain front velocities and accelerations, using Levenberg-Marquardt MPFIT routines \citep{Markwardt:2009} combined with a statistical bootstrapping technique \citep{Efron:1979} for measurement error minimization. The thick black solid line in Fig.\ref{fig_kinematics} shows the second-order fit we obtained. We found that the front velocity evolved between $449.1$~km/s and $282.8$~km/s, with deceleration of $-410$~m/s$^2$. We have assumed a radial expansion of the CBF, and have taken into account the geometrical foreshortening by multiplying heights above the limb by a factor of $1/\sin{(\phi)}$, $\phi$ being the heliospheric longitude of the eruption source. Behind the CBF, there is a distinct darkening, which may be a coronal dimming, and which we address in Section \ref{s3.1} below.

\subsection{Radio Observations}
\begin{figure}
\noindent\includegraphics[width=1.0\columnwidth]{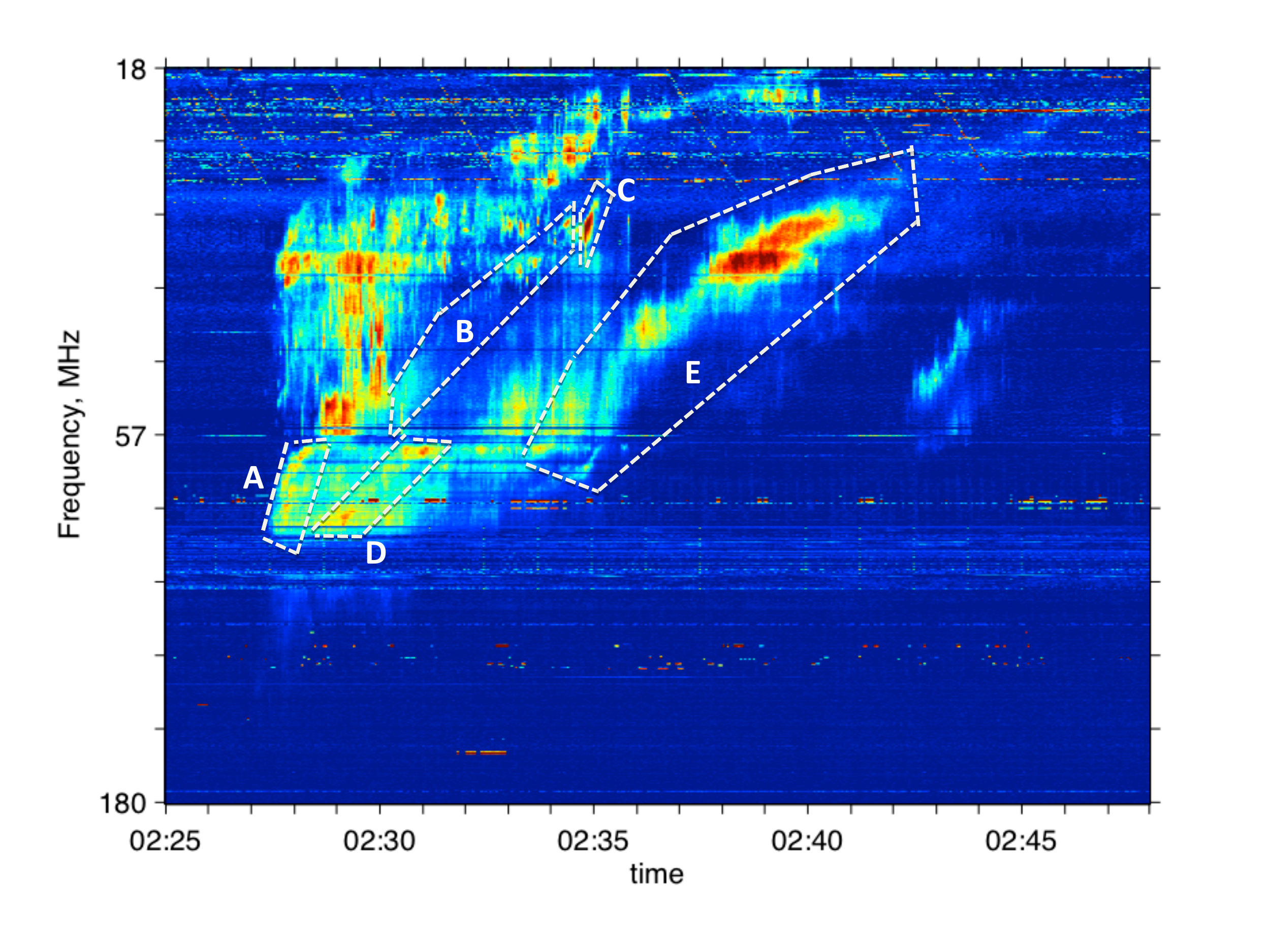}
\caption{A long duration metric radio type II burst was observed by the Culgoora radio spectrograph, between $\sim$02:27 and 02:46 UT, spanning the 18--150~MHz frequencies. The selected patches of the spectrogram have been used in the analysis (see Fig. \ref{fig_thbn_stats} and Section 3.2).}
\label{fig_radio}
\end{figure}

We show metric radio spectra during the period of the eruption from the Culgoora Solar Observatory (Fig. \ref{fig_radio}), operated by the Australian Radio and Space Weather Services. The instrument covers a frequency range 18Ð-1800 MHz and completes a frequency sweep every 3 s. The total frequency range is split into four bands. We use data from  the two lower bands, 18-75~MHz and 75-180~MHz. A distinct type II burst as well as type III bursts occurred during and after the EUV eruption, starting at 02:27 UT. The type II burst was weak at first, but eventually intensified. This observation confirms the presence of a shock wave during and after the EUV eruption.

\subsection{Particle Observations}
\begin{figure}
\noindent\includegraphics[width=0.8\columnwidth]{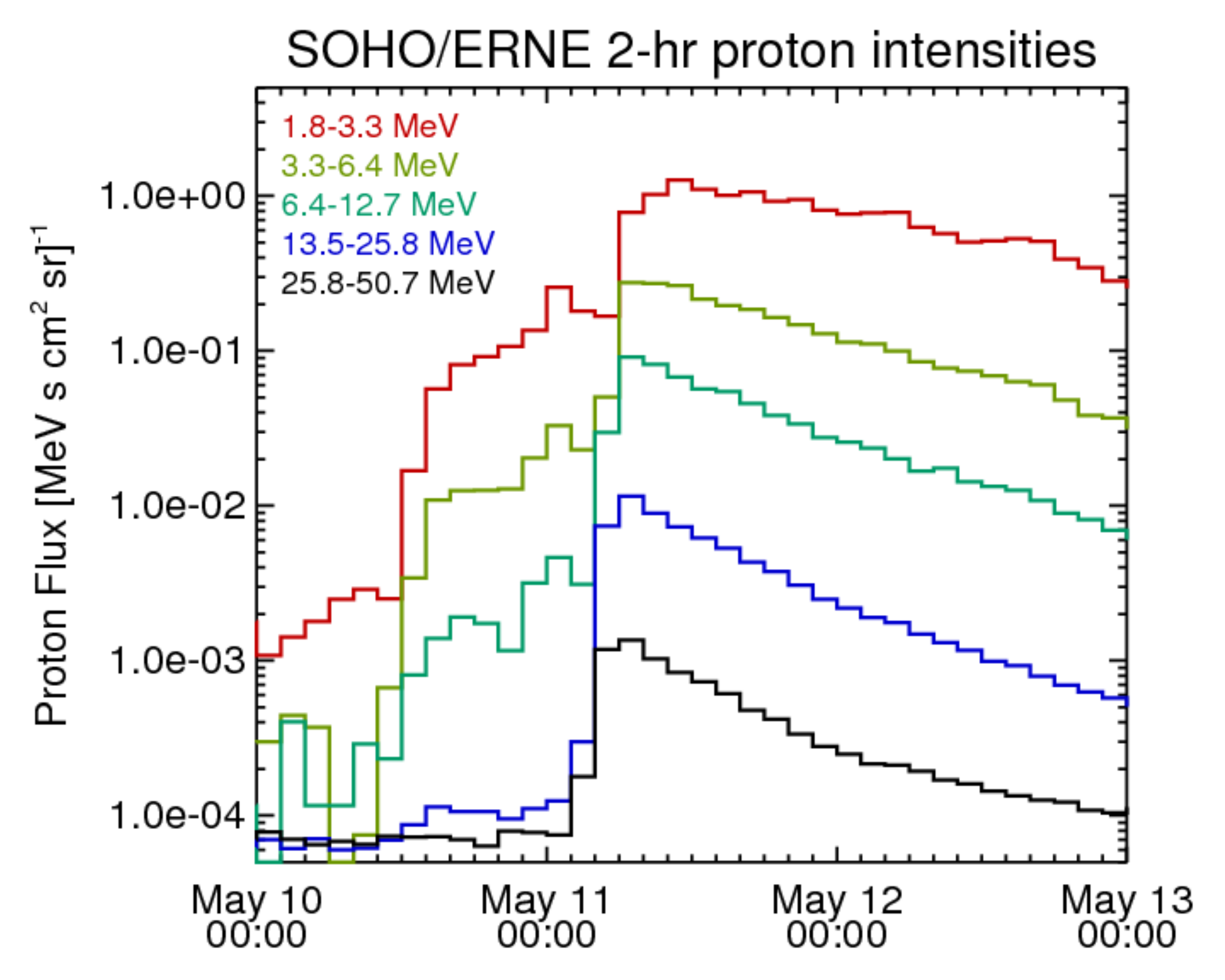}
\caption{Two-hour-averaged SOHO/ERNE observations of energetic proton differential fluxes at the L1 point for the three days surrounding the May 11 event. Five SEP channel time series are shown, in different colors. The minor ticks split the time axis in two-hour intervals. }
\label{fig_particles}
\end{figure}

Shortly after the solar onset of the eruption, a distinct increase in the proton fluxes was observed by SOHO/ERNE instrument at L1, at energies up to ~50 MeV (Figure \ref{fig_particles}). The May 11 event was ideally positioned to cause this increase in fluxes, as it occurred close to the western limb of the Sun and thus was likely very well magnetically connected to near-Earth space. Prior to the event, and unrelated to it, proton fluxes in the low-energy channels (1.8-12.7~MeV) were already elevated, but even in those channels the sharp increase can be easily seen. The fluxes in all but the lowest-energy channel peaked between 06 and 08 UT on May 11. In the lowest channel, the flux peaked some four hours later, which is consistent with longer travel times, as well as enhanced turbulence which may have been caused by the high-energy particles, reducing the scattering mean free paths in interplanetary space. The enhanced fluxes continued for more than a day after the event onset, in all channels.


\section{Low Corona Shock Acceleration Efficiency and Coronal Magnetic Connectivity}
\label{s3}

The coronal magnetic field configuration plays an important role in the production and release of energetic particle during explosive events in the solar atmosphere. In particular, we have attempted to assess the time-dependent orientation of the coronal magnetic fields to the wave/shock traveling between 1.2-1.5~\rsun, as well as its strength as judged by the change of emission measure. In the current theoretical framework of diffusive shock acceleration, a quasi-perpendicular orientation of the local shock surface normal direction to the magnetic field direction is favored for the fast acceleration of charged particles. In idealized particle simulations \citep{Giacalone:2006b}, perpendicular MHD shocks have been shown capable of accelerating ions and electrons to MeV energies fast enough to support coronal acceleration. At the same time, the number of particles accelerated at quasi-perpendicular shocks may depend on the suprathermal seed particle population, while parallel shocks have access to a larger fraction of the ambient particles but with slower energy gain \citep{Tylka:2005, Tylka:2006, Zank:2006}. As a first step to estimating SEP spectra from remote observations early in events, we have focused on obtaining time-dependent parameters related to shock strength and magnetic field orientation. To do that, we have combined AIA observations with several modeling tools.

\subsection{Shock Strength and Heating From Emission Measure Analysis}
\label{s3.1}
\begin{figure}
\noindent\includegraphics[width=1.0\columnwidth]{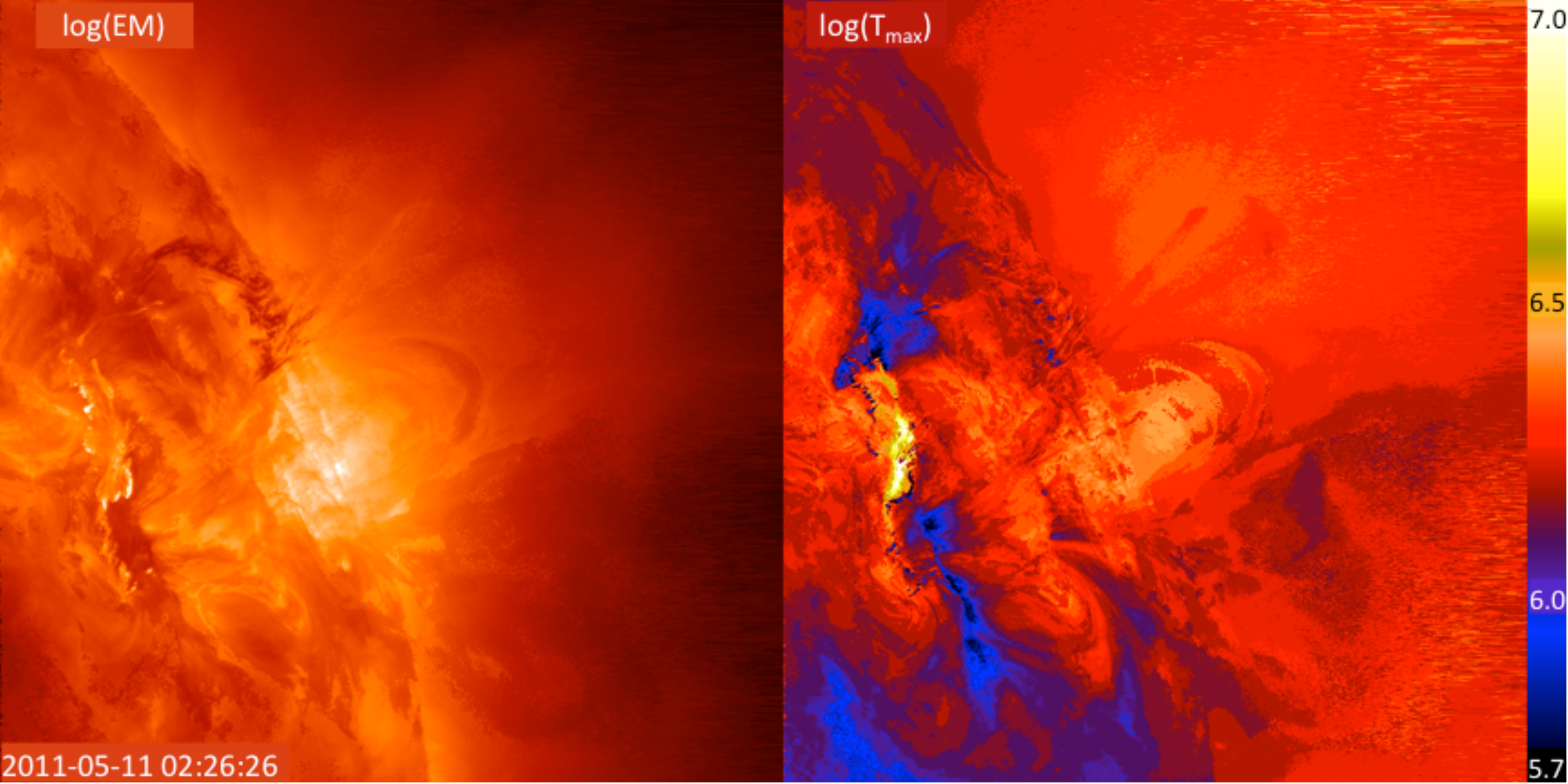}
\caption{Log of Emission measure ($EM$, left) and log of peak emission temperature ($T_{\max}$, right), obtained with the EM model of \citet{Aschwanden:2013} for a single time step during the event. The vertical legend is for the log($T_{\max}$) image on the right. A movie is available in the online version of the article.}
\label{fig_dem}
\end{figure}

In order to estimate the density change in the shock sheath, we used the DEM model of \citep{Aschwanden:2013}. It is written in IDL, and is included in the SolarSoft library package. The model uses the closest in time images (or sub-images) of the six EUV channels of AIA - 131, 171, 193, 211, 335, 94 $\AA$, combined with the wavelength response functions for each channel of the instrument, to calculate the observed DEM as a function of temperature. It fits a single Gaussian function to the DEM curve for each pixel of the image it receives as input, and outputs the temperature of maximum emission ($T_{\max}$), as well as the corresponding emission measure ($EM$). This is done separately for each set of six EUV-band images. The results of the model are presented here as a way to show the plasma compression (through the change in EM) and the heating, but we note that the heating may be underestimated, because the background emission dominates the intensities being fit. Using this model, we have built a time history of the temperature and emission measure in the AIA FOV, based on the EUV emission along the line of sight. Figure \ref{fig_dem} shows one snapshot of the model output. Since the CBFs are very dim relative to the solar disk, it is very hard to distinguish the shock sheath in one single snapshot, but it is easily seen in the movie, available in the online version of the article. 

\begin{figure}
\includegraphics[width=1.0\columnwidth,angle=0]{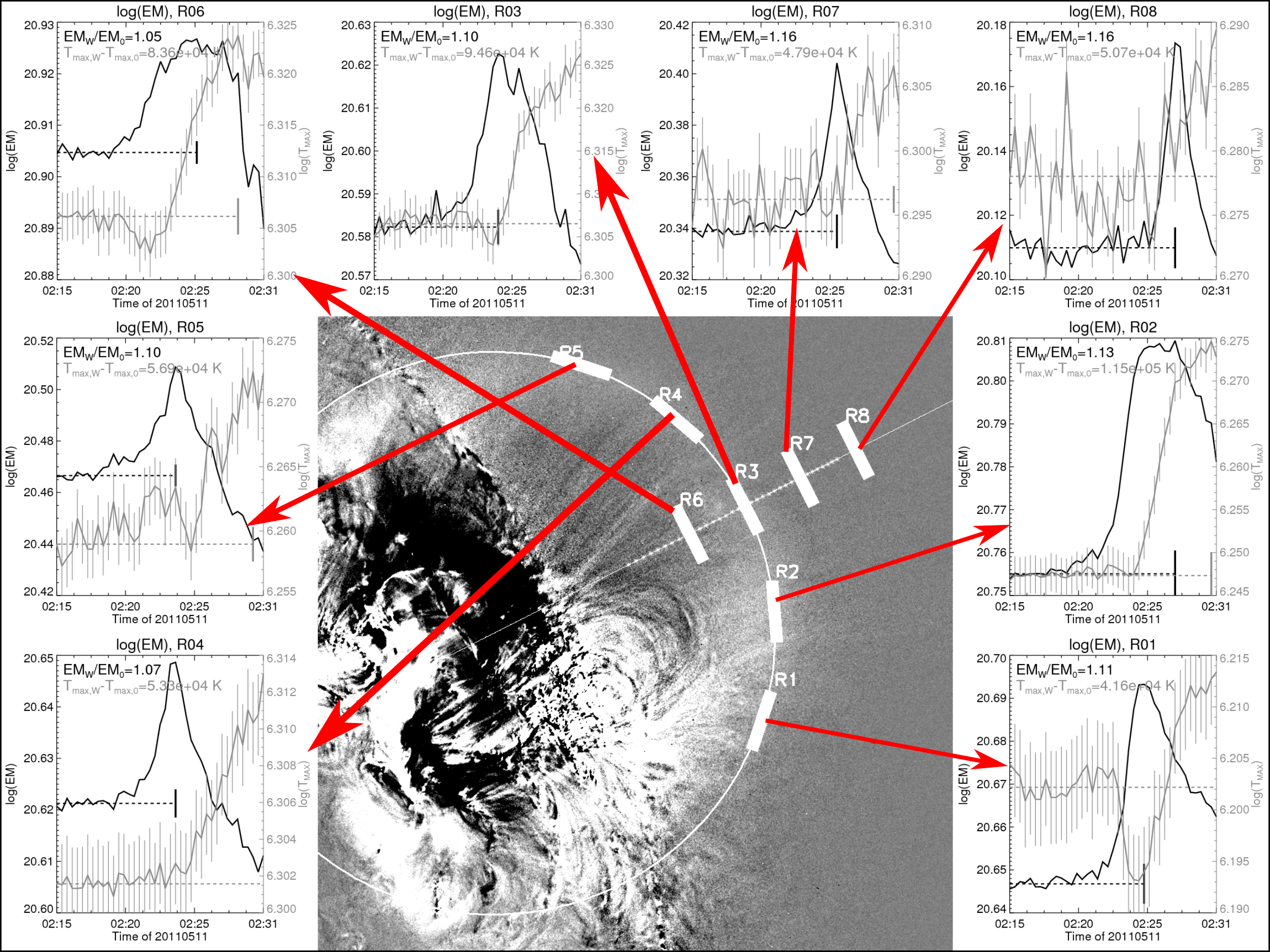}
\caption{Time evolution of $EM$ and $T_{\max}$ averaged for eight 200-pixel regions of the eruption. $EM$ is plotted in black, $T_{\max}$ in grey. The ratio of peak to background (dashed lines) $EM$ values for each region allows to estimate a density compression, assuming a constant emission volume \citep{Kozarev:2011}.}
\label{fig_density}
\end{figure}

To study the change in temperature and emission measure, we have selected eight 200-pixel regions (referred to as R1-R8) above the limb over the source location of the eruption (Figure \ref{fig_density}). For all time steps of the event, we calculated the average values of $T_{\max}$ and $EM$ for every region. The values are plotted as separate time series plots in Fig.\ref{fig_density}. On every one of those plots, $EM$ is plotted in black, while $T_{\max}$ is in grey. The vertical bars on the $T_{\max}$ curves denote the FWHM of the Gaussian fit to the temperature. The horizontal dashed black and grey lines mark the background values of $T_{\max}$ and $EM$, respectively, while the vertical bars at their ends mark the times when maximum values for the two quantities occurred. The peak times and values for all regions are given in Table \ref{table_demregions}. We note that varying the region size did not produce a noticeable difference in the time series of two quantities in question, except for the level of noise. The central image in the figure shows the locations of the regions on an AIA base difference image for context, with red arrows connecting each to its respective time series plot. In all regions, we see an immediate increase in the $EM$ after the passage of the shock front, peaking shortly after. The emission measure declined to significantly below the pre-eruption levels in all but the two southern regions (R1 and R2), possibly signifying a coronal dimming and plasma depletion behind the shock sheath. At the same time, there was an increase in $T_{\max}$, commencing around the $EM$ peak time. This would be consistent with a compressed sheath, followed by hotter but relatively rarified plasma passing through the selected regions. In a classical shock, one would expect the heating to peak very closely behind the shock surface, so this result is surprising. However, this separation between the EM and temperature peaks may occur because of one or both of the following: 1) a projection effect, whereby different position angles of the front enter the emitting column at different times; 2) an ionization effect - AIA effectively measures the ionization state of the plasma, and it can take several minutes for the ionization state to respond to a change of temperature at coronal densities.

\begin{table}[htc]
\centering
\begin{tabular}{c c c c c c c}
\hline
Region & $EM_{\rm peak}/EM_{\rm bckg}$ & t($EM_{\rm peak}$) & $T_{\rm peak}-T_{\rm bckg}$ & t($T_{\rm peak}$)\\
\hline
R1 & 1.11 & 02:25 & 4.16$\times10^4$ & 02:31\\
R2 & 1.13 & 02:26 & 1.15$\times10^5$ & 02:30\\
R3 & 1.10 & 02:24 & 9.46$\times10^4$ & 02:31\\
R4 & 1.07 & 02:24 & 5.38$\times10^4$ & 02:31\\
R5 & 1.10 & 02:24 & 5.69$\times10^4$ & 02:31\\
R6 & 1.05 & 02:25 & 8.36$\times10^4$ & 02:29\\
R7 & 1.16 & 02:25 & 4.79$\times10^4$ & 02:30\\
R8 & 1.16 & 02:27 & 5.07$\times10^4$ & 02:31\\
\hline
\end{tabular}
\caption{Peak to background emission measure ratios and temperature differences for the eight regions shown in Fig.\ref{fig_density}.}
\label{table_demregions}
\end{table}

We positioned regions R1, R2, R4, and R5 along the general curvature of the CBF, which allows to investigate the density and temperature behavior at different position angles along the CBF, but for similar structures (sheath, dimming). Comparing the T and EM time series for regions R1 and R2 to the south of the CBF nose, with those for regions R4 and R5, we see similar behavior - with the notable difference that there was no significant post-CBF coronal dimming at R1 and R2. The regions closer to the nose may have experienced more heating, but a more detailed study is necessary to confirm or reject such a conclusion.

The time series of regions R6, R3, R7, and R8 give us information about the heating and density change along the radial direction of propagation of the CBF. They are positioned in order of radial position on the top row in Fig. \ref{fig_density}. Comparing these four regions, we observe an increase, and consequent decrease of the T and EM change, with the largest EM ratio of all regions seen in R7. Again, we will explore the relations between position angle along the CBF and plasma sheath density and temperature changes in a future, more detailed study. Our results are consistent with the EM ratios of \citet{Kozarev:2011} and \citet[their Fig. 11]{Cheng:2012} - an increase of the EM in the sheath over the pre-event background between 3\% and 57\%. However, we note that in the case of 57\% increase the front was likely due to expanding magnetic loops rather than a shock sheath.

\subsection{Detailed Time-Dependent Shock Angles}
\label{shock_angles}
In order to estimate the shock acceleration efficiency, a very important parameter is the angle \thetabn, which helps determine the rate of momentum gained by particles, but changes rapidly as the shock propagates. In the following, we have combined a data-driven model of a three-dimensional spherical surface, propagating through the corona with a speed and radius based on the observed radial shock kinematics), which we call the Coronal Shock Geometric Surface (CSGS) model, together with a Potential Field Source Surface \cite[PFSS]{Schrijver:2003}) coronal magnetic field model, calculated for the period immediately preceding the event. 
\begin{figure}
\noindent\includegraphics[width=1.0\columnwidth]{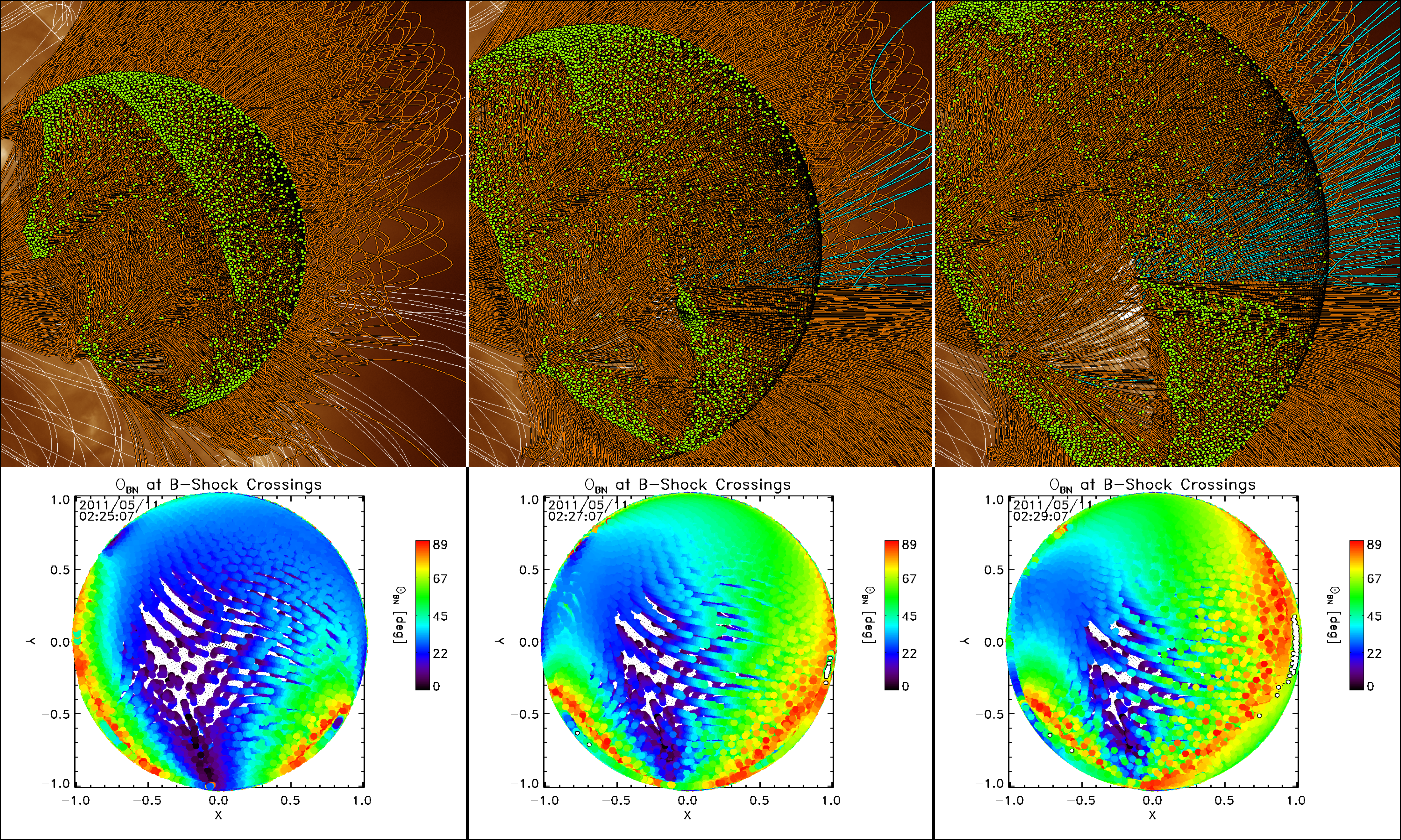}
\setlength{\abovecaptionskip}{-1pt}
\caption{{\bf Top row:} Three snapshots of the time-dependent coupled PFSS+CSGS model, showing the interaction of the spherical geometric shock front model with the PFSS coronal fields. The AIA 193-Angstrom channel image is shown for reference, for each time step. On top of it we have plotted the field lines - colored orange (closed) or blue (open) if interacting with the shock surface, white otherwise. The shock surface mesh is plotted in green. The orange dots represent the points of interaction. A full-resolution magnetic field movie is available in the online version of this manuscript. {\bf Bottom row:} For each step of the shock evolution, we produced a map of the position of the field interaction with the shock. The colors correspond to the value of the angle \thetabn~between 0\degrees~and 90\degrees. Open field-crossing symbols are open circles (their centers are white).}
\label{fig_shock}
\end{figure}

The PFSS model is freely available in the IDL SolarSoft package, and can be run on demand, in both interactive and batch mode. Full-disk, line-of-sight magnetogram data from SDO/HMI are assimilated into an evolving radial flux dispersal model, which is continuously sampled every six hours, and saved into maps. The PFSS model uses these photospheric magnetic maps as lower boundary conditions for field extrapolation, providing a global coronal vector magnetic field solution for a 3D grid of polar coordinates. The self-consistent evolving flux model mitigates issues with the PFSS model performance near the solar limb, where magnetogram data is not accurate. To ensure that the model used for the May 11 event is appropriate, we have visually inspected the maps by overlaying them on AIA 171 A images, and found very good agreement between active region positions and the origins of magnetic lines. For this application, the model was run in batch mode, and calculated the extrapolated coronal field lines with starting points inside a region of $90\times90$\degrees~around the source location of the eruption. The lower and upper boundaries were set to 1.05 and 2.5~\rsun, respectively.

The CSGS model is written in IDL. It takes as input the fitted time-dependent position of the shock front in the radial direction, and produces a three-dimensional spherical dome (or cap) surface, using IDL's surface of revolution and 3D visualization functions. The radius of the dome is taken as the distance between the shock front nose and the eruption source location (that is, the center of the spherical dome is fixed). At any given time step, the points of intersection between the current CSGS surface and any PFSS field line that crosses it, are found. Then, the local normal to the shock surface is calculated, and with it, the upstream angle \thetabn. All the information is saved, and the model proceeds to the next time step. CSGS keeps a record of which field lines interacted with the shock for how long, making it ideal to use with models for particle acceleration. We are in the process of extending the model to non-spherical geometries.

Figure \ref{fig_shock} shows the application of the combined PFSS-CSGS models. The top row shows the AIA/193 images at three time steps, separated by two minutes from each other. The solar limb is also shown with a white line. Overlaid are the PFSS field lines and the CSGS model. The field lines interacting with the shock surface are colored, while the ones not interacting are white. Closed lines are in orange, while open lines are in light blue). The shock surface mesh is in black. The points of interaction are shown in light green. From the sequence it can be seen that the number of interacting field lines increases with time, and that the points of interaction `slide' along the shock surface. The model is applied for every time step of AIA data used for this analysis.

The bottom row shows the calculated \thetabn~angles as a function of position around the shock surface for the same three time steps as the top row. Each plot contains all points of interaction accumulated up to that time step. Their positions have been scaled to a unit spherical dome surface, in order to be able to compare them across time steps. The nose of the shock is in the center of the plot. The surface has not been rotated, i.e., North is up, and West is right. The colors correspond to \thetabn~values between 0 and 90\degrees. Comparing the three plots, it is immediately visible that the field-shock angle changes significantly throughout the evolution of the shock surface, and that areas of higher \thetabn~values, which correlate with faster and stronger acceleration, occur preferentially near the flanks for the magnetic fields calculated by PFSS for this event. In addition, we have shown the crossings of open field lines with open circular symbols, keeping the color coding. These can be seen near the right (western) edge of the shock in the second and third panels, corresponding to the light blue field lines in the top row panels. The open field line crossings occur near the high-\thetabn~(orange and red) angle region of the shock surface.

\begin{figure}
\includegraphics[width=1.0\columnwidth]{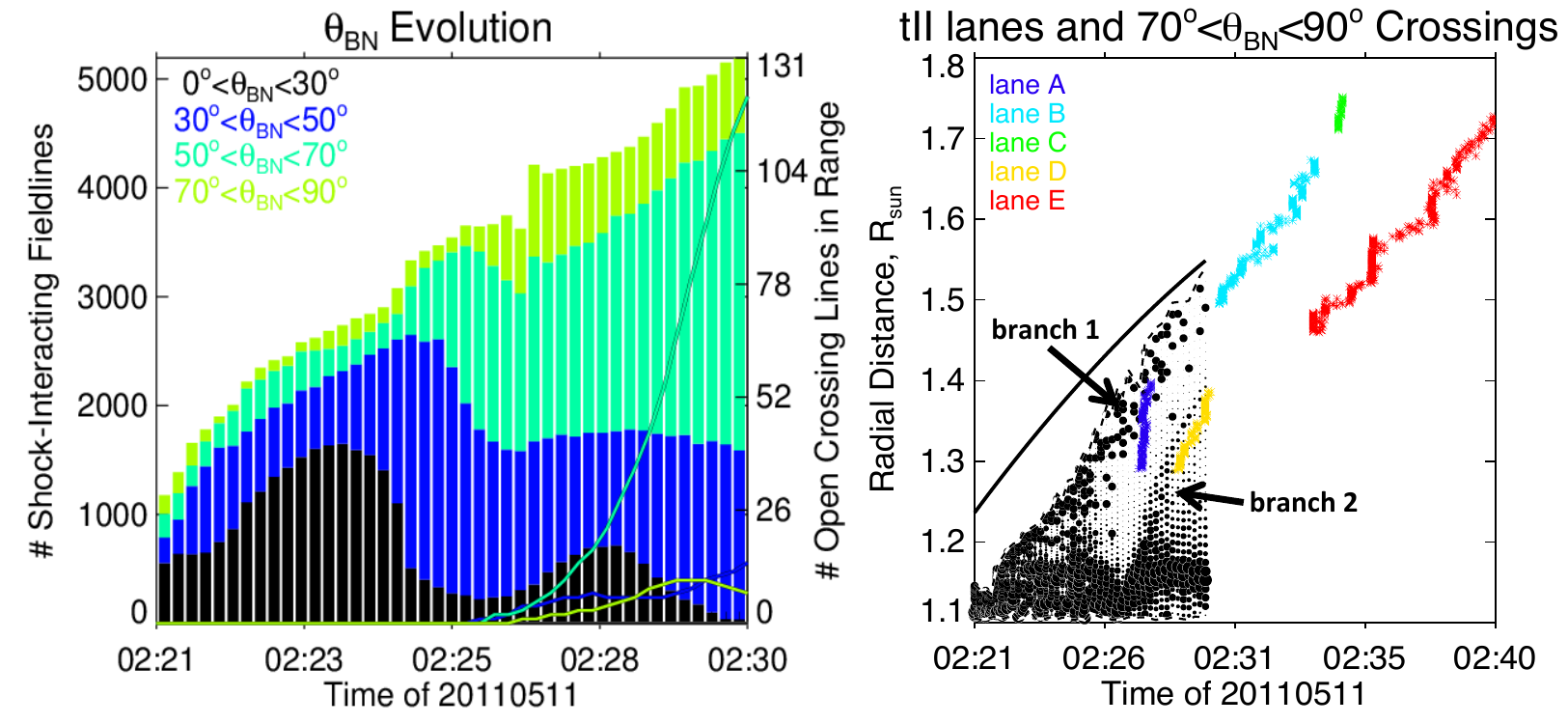}
\setlength{\abovecaptionskip}{-1pt}
\caption{\textbf{Left:} A distribution of the calculated shock-field angles \thetabn, for four angle intervals, as a function of the shock nose height in the AIA FOV. The number of open field lines crossed by the shock as a function of time in each angle interval is also plotted as lines with corresponding colors. \textbf{Right:} Time series of the crossing radial heights for the 70\degrees$<$\thetabn$<$90\degrees range from the left panel (black filled circles); minimum and maximum heights are shown with dashed lines. Type II emission maxima heights selected along the lanes marked in Fig. \ref{fig_radio}, based on a one-fold Newkirk coronal density model \citep{Newkirk:1961}, are shown with colored asterisks. The thick solid line is the fitted shock nose position.}
\label{fig_thbn_stats}
\end{figure}

To explore in more detail the orientation of the PFSS field lines with respect to the model shock surface, we gathered statistics on \thetabn~as a function of radial distance from the Sun. Figure \ref{fig_thbn_stats}, left panel, shows a summary of the \thetabn~values over the duration of the event in the AIA FOV, which was fitted with the CSGS model. We show four angle intervals, broadly corresponding to quasi-parallel (0-50\degrees), and quasi-perpendicular (50-90\degrees) \thetabn~angles. In addition, we have overplotted the number of open field line crossings in each angle range as line time series, with values marked on the right side of the panel. For this specific event, we observe that: 1) the number of interacting field lines more than doubles over the duration of the event, as the shock surface becomes larger; 2) whereas in the beginning of the event the shock is more quasi-parallel, in the second half of the event the quasi-perpendicular angles are a much bigger fraction of the overall interaction; c) the number of open field lines crossing the shock increases significantly after the . We note that in diffusive shock acceleration theory, acceleration efficiency does not depend linearly on \thetabn, but is proportional to $1/\cos($\thetabn$)^2$ (assuming negligible perpendicular diffusion). Thus, the shock is expected to increase its acceleration efficiency significantly in the second portion of its low corona evolution. 

The right panel of Fig. \ref{fig_thbn_stats} shows time series of the heights of all crossings in the 70\degrees$<$\thetabn$<$90\degrees range (filled black circles, with dashed lines denoting the minima and maxima). At every time step we have binned the crossings into 40 bins of radial height, with circle sizes normalized to the largest bin value at that time step. Thus, most of the model crossings in this degree range occur quite low in the corona throughout the event; however, we see two distinct `branches' of crossings - branch 1, with relatively many crossings per time step, starts relatively early in the event and brings up the largest crossing heights ever closer to the fitted shock nose (thick solid line); a smaller (fewer crossings per time step) branch (branch 2) starts low around 02:26 UT, and reaches almost 1.4\rsun by the end of the event.

On the same panel, we have overplotted the radial heights corresponding to radio type II burst emission lanes (colored asterisks), calculated using a one-fold Newkirk density model \citep{Newkirk:1961}. The selected bands are shown on Fig. \ref{fig_radio}. Upon visual inspection, we determined that all but lane C are due to harmonic emission. We find that the two main bands of radio emission heights correspond roughly to the two branches of the shock crossing heights, and the emission continues beyond the time range of the AIA data. In fact, we find that lane D is spatially very close to branch 2, and has the same slope, while lane A overlaps with part of the branch 1, but has a slightly steeper slope. This results validates our model and supports electron acceleration near shock crossings of large \thetabn, as expected in DSA theory. We will explore this connection further in future work through studying multiple events.

\begin{figure}
\includegraphics[width=1.0\columnwidth]{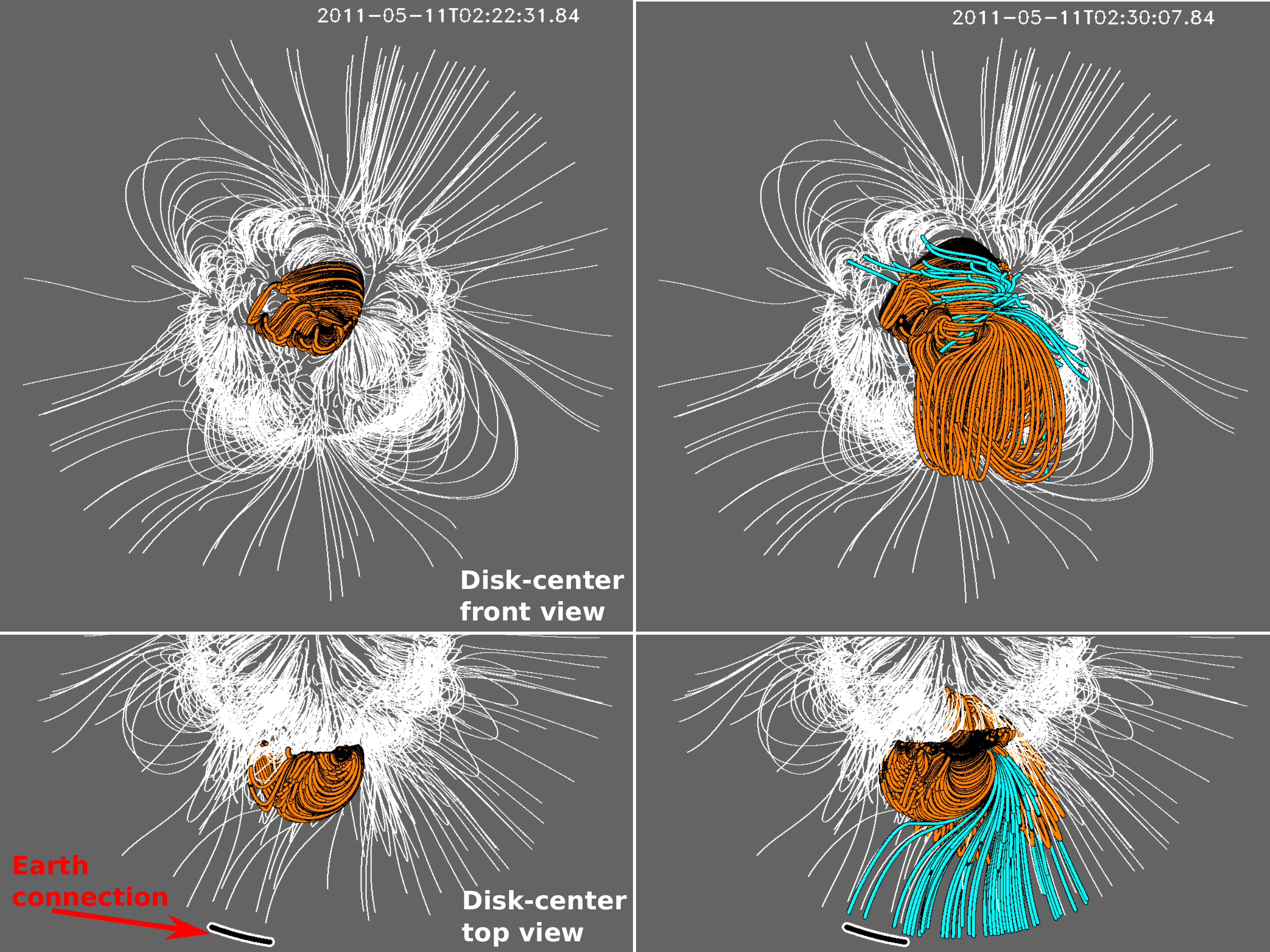}
\setlength{\abovecaptionskip}{-1pt}
\caption{Two snapshots of the PFSS+CSGS model show the evolution of heliospheric connectivity of the front-crossing coronal field lines. Orange field lines are closed, blue lines are open. The left and right panels correspond to the initial and final step measured in the AIA FOV for this event, respectively. The region of the shock is shown at disk center in the top panels. In the bottom panels a polar (top-down) view is shown. The black arcs near the bottom show the connectivity to the Earth along a Parker spiral magnetic field corresponding to solar wind speeds between 400 and 500 km/s.}
\label{fig_spread}
\end{figure}

\subsection{Heliospheric Magnetic Connectivity of the Event}
The method to determine the time-dependent \thetabn~may also be used to estimate the heliospheric magnetic connectivity of open coronal field lines during the event. Figure \ref{fig_spread} shows the result of applying the CSGS model on a low-density PFSS model. The left and right columns correspond to the first and last measured shock nose positions in the AIA FOV (separated by about eight minutes). We have de-rotated the PFSS+CSGS model, and chosen a viewpoint that puts the eruption source location at disk center. The top panels show the model face-on, while the bottom panels show a polar view. The field lines interacting with the model shock surface have been colored for distinction. The small black arc at the bottom of those panels represents the range of nominal magnetic connections to near-Earth space, assuming a Parker spiral interplanetary magnetic field and solar wind speed ranging between 400 and 500~km/s. It can be seen that at the last time step of this model, there was already solid nominal connection to near-Earth space. It is also interesting to note that the connected open field lines originate near the west side of the shock surface, and wrap around it in a way that enhances the quasi-perpendicularity of the shock near its nose. Overall, this approach holds much promise for future predictive capability, although we will test its validity on multiple events first.


\section{Summary}
\label{s4}

We have analyzed the May 11, 2011 solar CBF, observed in SDO/AIA EUV images, and determined it to be the signature of a shock wave, driven by an erupting filament, and accompanied by the prompt arrival of SEPs at L1, as well as a distinct metric type II radio burst. We found that the radial speeds of the shock front evolved between about 450 and 280~km/s in the AIA field of view. We modeled the time-dependent emission measure and temperature of the wave using a recent DEM model. We found: a) a weak density enhancement at several different positions along the front; b) The maximum emission temperature increased after the passage of the wave front, while the emission measure decreased below pre-event values, consistent with a shock sheath followed by a hot coronal dimming. 

We used the shock kinematics measurements to fit a spherical geometric surface (CSGS model) to the shock for every time step we could measure. We calculated a PFSS coronal field model for the period preceding the event, and combined it with the CSGS model, in order to determine time-dependent shock-crossing angles \thetabn. We found that:  a) the \thetabn~values during this event were predominantly less than 50\degrees in the first half of the event, while values closer to 90\degrees dominated the second half; b) for the domain over which the event was observed (1.1--1.5\rsun), the shock was quasi-parallel near its nose in the beginning of the event, but turned more quasi-perpendicular later on, as well as near its flanks throughout the event; c) The model for the time-dependent radial heights of the field-shock crossings shows remarkable consistency with the radial heights of the type II burst emission determined from a simple coronal density model; d) the open field lines (from the PFSS model), which the shock crossed near its western edge in the second half of the event, were likely both related to high \thetabn~angle values, and magnetically connected to near-Earth space. Based on these results, the shock acceleration efficiency and particle release should be considerably higher in the second part of the shock evolution in the AIA FOV. In future work, we will test the relationship between the evolution of these parameters and the particle fluxes (both in the corona and near Earth) by combining the tools presented here with a shock acceleration model. This will allow us to quantify the production and release of energetic particles in the corona during real events, and subsequently focus on their interplanetary propagation and the validation of the model results.

The analysis of the observations presented here suggests that the coronal magnetic geometry is very important for particle production and release in explosive events. We used these observations to drive several models that provide an insight into the shock thermal and magnetic properties: using a DEM model to obtain information about how coronal plasma density and temperature changed as a function of time and position around the shock; combining a PFSS coronal field model with the CSGS coronal shock model we built, we were able to estimate the \thetabn~shock-to-field angle evolution. While the validity of our results is limited by the validity of the PFSS and DEM models we used, these techniques nevertheless reveal valuable information about the low coronal evolution and interaction of shocks with overlying magnetic fields and coronal plasma. In future work, we will apply the methods introduced here to studying multiple events observed with AIA.


\acknowledgments
We acknowledge support under AIA subcontract SP02H1701R from Lockheed-Martin. KAK was supported under the NASA Living With a Star Jack Eddy Postdoctoral Fellowship Program, administered by the UCAR Visiting Scientist Programs.


\end{document}